\documentclass[twocolumn,showkeys,
  preprintnumbers,amsmath,amssymb]{revtex4}

\newcommand{\nix}[1]{{}}
\newcommand{\floor}[1]{{\lfloor#1\rfloor}}
\newtheorem{theorem}{Theorem}

\begin{document}
\preprint{}

\title{Asymptotics of the Quantum Hamming Bound for Subsystem Codes}

\author{
Andreas Klappenecker
and 
Pradeep Kiran Sarvepalli
}
\affiliation{
Department of Computer Science,
Texas A\&M University,
College Station, TX 77843, USA
}

\date{\today}

\begin{abstract}
Ashikhmin and Litsyn showed that all binary stabilizer codes -- pure
or impure -- of sufficiently large length obey the quantum Hamming
bound, ruling out the possibility that impure codes of large length
can outperform pure codes with respect to sphere packing. In contrast
we show that impure subsystem codes do not obey the quantum Hamming
bound for pure subsystem codes, not even asymptotically.  We show that
there exist arbitrarily long Bacon-Shor codes that violate the quantum
Hamming bound.
\end{abstract}

%\pacs{03.65.Nk, 03.65.Pm, 24.10.Jv}
\keywords{quantum Hamming bound, asymptotic bounds, degenerate codes,
subsystem codes} \maketitle
%\section{Introduction}
Degenerate quantum error-correcting codes pose many interesting
questions in the theory of quantum error-correction. The early
discovery of the phenomenon of degeneracy raised the question whether
degenerate quantum codes can perform better than nondegenerate quantum
codes. One of the unresolved questions to this day in the theory of
stabilizer codes is whether the bounds that hold for nondegenerate
codes also hold for degenerate codes. Some bounds like the quantum
Singleton bound do. But for others, like quantum Hamming bound, an
answer remains elusive. Partial answers were provided by Gottesman
\cite{gottesman97} for single error-correcting and double
error-correcting codes. Ashikhmin and Litsyn \cite{ashikhmin99} showed
that asymptotically degenerate codes cannot beat the quantum Hamming
bound. \textit{This leaves only a small range of degenerate binary stabilizer
codes of moderate length that can potentially beat the quantum Hamming
bound, but we conjecture that no such examples can be found. }

We show that the situation is markedly different in the case of
subsystem codes (also known as operator quantum error-correcting codes
\cite{kribs05,knill06,kribs06}). The quantum Hamming for pure
subsystem codes was derived in \cite{aly06}. In \cite{pre0703}, it was
shown that there exist impure subystem codes that beat the quantum
Hamming bound for pure subsystem codes. However, it remained unclear
whether impure subsystem codes asymptotically obey the quantum Hamming
bound, as in the case of binary stabilizer codes. The purpose of this
note is to show that there exist impure subsystem codes of arbitrarily
large length that beat the quantum Hamming (or sphere-packing) bound.

Recall that the quantum Hamming bound for subystem codes states that a
pure $[[n,k,r,d]]$ subsystem code satisfies
\begin{eqnarray}\label{hamming} 
2^{n-k-r}	\geq \sum_{j=0}^{\floor{(d-1)/2}}\binom{n}{j}3^j.\label{eq:qhb}
\end{eqnarray}
For all positive integers $n$, there exist subsystem codes with
parameters $[[n^2,1,(n-1)^2,n]]$ -- the Bacon-Shor codes, see \cite{bacon06a,bacon06b}.  
We claim that all $[[(2t+1)^2,1,4t^2,2t+1]]$ subsystem codes violate the
quantum Hamming bound, namely that
\begin{eqnarray*}
2^{(2t+1)^2-1-4t^2}=	2^{4t}&\not\geq &\sum_{j=0}^{t}\binom{(2t+1)^2}{j}3^j
\end{eqnarray*}
holds for all positive integers $t$. It suffices to show that 
\begin{eqnarray}\label{ineq}
2^{4t}&< &\binom{(2t+1)^2}{t}3^t
\end{eqnarray}
holds for all positive integers $t$. Since $0<4(t-1/6)^2+8/9=4t^2-4t/3+1$, we have 
$$ \frac{16t}{3} <{4t^2+1+4t}$$ for all $t>0$. Multiplying both sides
by $3/t$ and raising to the $t^{th}$ power yields
$$2^{4t}< \frac{3^t(2t+1)^{2t}}{t^t}, $$ which proves the inequality
(\ref{ineq}), as $\binom{n}{k} \geq n^tk^{-t}$.  Thus, we can conclude
that the Bacon-Shor codes of odd length do not obey the quantum
Hamming bound.

\begin{theorem}
Asymptotically, the quantum Hamming bound~(\ref{hamming}) does not
hold for impure subsystem codes.
\end{theorem}

It is remarkable that there exist such families of subsystem codes
that can pack more densely than any pure subsystem code. Further
examples of such densely packing subsystem codes can be found among
the family with parameters 
$[[n_1n_2,1,(n_1-1)(n_2-1),\min\{n_1,n_2\}]]$, which contains for
instance a $[[12,1,6,3]]$ subsystem code.

\def\cprime{$'$}

%\bibliographystyle{plain}
%\bibliography{refs} 
\end{document}